\documentclass[journal]{IEEEtran}
\usepackage{amsmath,amsfonts}
\usepackage{algorithmic}
\usepackage{algorithm}
\usepackage{array}
\usepackage[caption=false,font=normalsize,labelfont=sf,textfont=sf]{subfig}
\usepackage{textcomp}
\usepackage{stfloats}
\usepackage{url}
\usepackage{bm}
\usepackage{verbatim}
\usepackage{graphicx}
\usepackage{booktabs}
\usepackage{multirow}
\usepackage{cite}
\usepackage[nolist]{acronym}
\usepackage[implicit=false]{hyperref}
\hypersetup{hidelinks,
	colorlinks=true,
	allcolors=black,
	pdfstartview=Fit,
	breaklinks=true}

\newcommand{\bl}[1]{{\color{black}#1}}

\begin{document}

\title{GLDPC-PC Codes: Channel Coding Towards\\ 6G Communications}

\author{Li Shen,
		~Yongpeng Wu,
		~Yin Xu,
		~Xiaohu You,
		~Xiqi Gao,
		~and Wenjun Zhang
\thanks{Li Shen, Yongpeng Wu, Yin Xu, and Wenjun Zhang are with Shanghai Jiao Tong University; Xiaohu You and Xiqi Gao are with Southeast University.}
}

\begin{acronym}
	\acro{2G}{second generation}
	\acro{3G}{third generation}
	\acro{4G}{fourth generation}
	\acro{5G}{fifth generation}
	\acro{6G}{sixth generation}
	\acro{AM}{adjacency matrix}
	\acro{APP}{a-posteriori probability}
	\acrodefplural{APP}{a-posteriori probabilities}
	\acro{AWGN}{additive white Gaussian noise}
	\acro{BCH}{Bose–Chaudhuri–Hocquenghem}
	\acro{BCJR}{Bahl-Cocke-Jelinek-Raviv}
	\acro{BEC}{binary erasure channel}
	\acro{BLER}{block error rate}
	\acro{BP}{belief propagation}
	\acro{B-DMC}{binary-input discrete memoryless channel}
	\acro{CN}{check node}
	\acro{CRC}{cyclic redundancy check}
	\acro{GLDPC}{generalized LDPC}
	\acro{GLDPC-PC}{GLDPC with polar-like component}
	\acro{GM}{generation matrix}
	\acro{i.i.d.}{independent and identically distributed}
	\acro{LDPC}{low-density parity-check}
	\acro{LLR}{log-likelihood ratio}
	\acro{MAP}{maximum a posteriori}
	\acro{MIMO}{multiple-input multiple-output}
	\acro{MPA}{message passing algorithm}
	\acro{NB-LDPC}{non-binary LDPC}
	\acro{PAC}{polarization-adjusted convolutional}
	\acro{PCM}{parity check matrix}
	\acro{RCU}{random coding union}
	\acro{SC}{successive cancellation}
	\acro{SCL}{successive cancellation list}
	\acro{SC-LDPC}{spatially coupled LDPC}
	\acro{SISO}{soft-input soft-output}
	\acro{SO-GRAND}{soft-output guessing random additive noise decoding}
	\acro{SO-SCL}{soft-output SCL}
	\acro{SNR}{signal-to-noise ratio}
	\acro{VN}{variable node}
\end{acronym}

\maketitle

\pagestyle{empty}
\thispagestyle{empty}

\begin{abstract}
The sixth generation (6G) wireless communication system will improve the key technical indicators by one to two orders of magnitude, and come with some new features. As a crucial technique to enhance the reliability and efficiency of data transmission, the next-generation channel coding is thus confronted with new challenges in terms of complexity, latency, performance. This article supplies an overview of the potential channel codes for 6G communications. In addition, we explore to develop next-generation channel codes based on low-density parity-check (LDPC) and polar frameworks, introducing a concept named generalized LDPC with polar-like component (GLDPC-PC) codes. The codes have exhibited promising error correction performance and manageable complexity, which can be further optimized by specific code design. The opportunities and challenges of GLDPC-PC codes are also discussed.
\end{abstract}


\section{Introduction}
In communication systems, the received signal may be corrupted by various factors, like noise, interference, and signal attenuation, leading to unexpected errors during data transmission. Channel coding technique combats these errors and enhances the reliability and efficiency of transmission by introducing additional redundancy. Owing to this, channel coding is essential in many modern communication systems, e.g. wireless communication, satellite communication, digital broadcasting, and fiber-optic communication systems.

Since Shannon's channel coding theorem has demonstrated that, by employing appropriate channel coding techniques, the transmission rate can approach the channel capacity with arbitrarily low error probability, numerous researchers have been dedicated to designing effective channel coding schemes. In the 1950s, the introduction of convolutional codes marked a significant milestone in channel coding development. Specially, the invention of the Viterbi decoding algorithm greatly promoted the practical application of convolutional codes, and emphasized the importance of soft information inputs for the channel decoder. In the 1990s, the emergence of turbo codes revolutionized the field of channel coding. Turbo codes, introduced by Claude Berrou \emph{et al.}, demonstrated exceptional error correction performance close to the theoretical limits predicted by Shannon. The implementation of iterative turbo decoding also inspired the rediscovery of \ac{LDPC} codes \cite{Gallager1962Low} in the early 2000s, which were initially proposed by Gallager in the 1960s and were ignored for a long time because their decoding complexity was unacceptable at that time. Using the advanced \ac{BP} decoding algorithm, \ac{LDPC} codes have demonstrated excellent performance that is extremely close to channel capacity on many channels. In the 2009, the polar codes \cite{Arikan2009Channel}, proposed by Arıkan, have emerged as a groundbreaking development in channel coding. Polar codes are based on the theory of channel polarization and are the first coding method with a deterministic construction that has been proved to achieve the capacity of symmetric \acp{B-DMC}.


For future \ac{6G} mobile communication systems, the key performance indicators, e.g. peak data rate, latency, and reliability, will be improved by one or two magnitudes compared to \ac{5G} systems \cite{Rowshan2024Channel}. Hence, we need to further develop the next-generation channel codes to address the stringent requirements of \ac{6G}. First, lower complexity, lower latency, and performance closer to the Shannon limit are unavoidable topics in channel coding design. Moreover, an increasing number of delay-sensitive services have posed serious challenges to the design of short codes, especially in terms of error correction performance and ultra-low latency decoders. Considering the potential integration of sensing with communication in \ac{6G}, the next-generation channel codes are required to balance the performance of these two purposes. Furthermore, a unified coding architecture that can exhibit excellent performance in a wide range of code lengths and code rates is also welcomed. This will bring various benefits, such as simplified system design, enhanced compatibility, and improved area efficiency.

In this article, we first review some state-of-the-art channel codes, including \ac{LDPC} codes \cite{Gallager1962Low}, \ac{GLDPC} codes \cite{Tanner1981Recur}, \ac{NB-LDPC} codes \cite{Davey1998Low}, \ac{SC-LDPC} codes \cite{Kudekar2011Thres}, polar codes \cite{Arikan2009Channel}, and \ac{PAC} codes \cite{Arikan2019From}. Then, we explore the possibility of developing next-generation channel codes on top of \ac{LDPC} and polar frameworks, and introduce a concept for channel codes towards \ac{6G}, named \ac{GLDPC-PC} codes. \bl{By leveraging the advanced soft-output \ac{SCL} decoder \cite{Yuan2024Near} for \ac{GLDPC-PC} decoding, they exhibit immense potential for ultra-reliable communications}. The opportunities and challenges of \ac{GLDPC-PC} codes are also discussed.

\section{State-of-the-art Channel Codes}
\subsection{LDPC Codes}
\Ac{LDPC} codes are a class of linear block codes with a sparse \ac{PCM}, which has a low density of 1's in the matrix. In addition to the matrix representation, the Tanner graph \cite{Tanner1981Recur}, a bipartite graph, is usually used to represent \ac{LDPC} codes. For a systematic \ac{LDPC} code, a codeword consists of the information bits and the remaining single parity checks. Each row of the \ac{PCM} corresponds to a parity check, named \ac{CN} in the Tanner graph, while each column of the \ac{PCM} corresponds to an encoded bit in the codeword, named \ac{VN}. When the element at the i-th row and the j-th column of the \ac{PCM} is a 1, the i-th \ac{CN} is connected to the j-th \ac{VN} in the Tanner graph. A single parity check is then said to be satisfied if and only if the bits at \acp{VN} that are connected to this \ac{CN} sum, modulo two, to zero. 

The encoding process of LDPC codes involves multiplying the message sequence by the \ac{GM} to generate the codeword, which can be obtained from the \ac{PCM} by ensuring that the parity checks hold for any valid codeword. At the receiver, the \ac{LDPC} decoder detects and corrects errors in the received noisy codeword under the parity check constraints. A well-known decoding algorithm is the \ac{MPA}, which refers to a class of iterative decoding algorithms, including the \ac{BP} algorithm and its approximations. The algorithm iteratively exchanges soft information between \acp{VN} and \acp{CN} along the edges in the Tanner graph. Due to their remarkable performance, various wireless communication standards, such as WiMAX, ATSC 3.0, DVB S2/T2, Wi-Fi, and 5G, have adopted \ac{LDPC} codes.

\subsection{GLDPC Codes}
As an extension of \ac{LDPC} codes, \ac{GLDPC} codes \cite{Tanner1981Recur} were first introduced by Tanner in 1981. Compared to conventional \ac{LDPC} codes, \ac{GLDPC} codes have a more flexible structure. In \ac{GLDPC} codes, the \acp{CN} in the Tanner graph are constrained to be a linear block code, rather than the single parity-check code. The sub-codes associated with the \acp{CN} are referred to as \textit{component codes}. Various component codes have been considered in \ac{GLDPC} code research, including Hamming codes, Bose–Chaudhuri–Hocquenghem codes, convolutional codes, etc. Polar codes have also been investigated as component codes for product codes \cite{Coskun2024Prec}. \bl{Product codes constrain the rows and columns of information bits in matrix form to be two linear block codes, while the \ac{CN} constraints of \ac{GLDPC} codes are more liberal. Hence, product codes can be regarded as a special structure of \ac{GLDPC} codes.}

Since \ac{GLDPC} codes share a similar code structure with LDPC codes, the \ac{MPA}, commonly used for \ac{LDPC} decoding, is also applicable to \ac{GLDPC} decoding. During the iterative decoding process, the soft information passed from \acp{CN} to the \acp{VN} can be obtained by performing a soft-output decoder of the component codes. 
Owing to the stronger constraints at the \acp{CN}, \ac{GLDPC} codes have demonstrated improved decoding performance, faster iterative convergence speed, and lower error floors for some specific code lengths at moderate-to-high \acp{SNR}. Hence, it makes \ac{GLDPC} codes a candidate for future wireless communication scenarios that require ultra-high reliability.

\subsection{NB-LDPC codes}
\ac{NB-LDPC} codes are an extension of \ac{LDPC} codes from the binary domain to higher-order Galois fields \cite{Davey1998Low}. That is, both the \acp{PCM} and codewords of \ac{NB-LDPC} codes take values in the Galois field, and the parity checks are also defined over the Galois field. \ac{NB-LDPC} codes can also perform \ac{BP} decoding in the Galois field, i.e., the q-ary sum-product algorithm. However, the complexity of the algorithm increases with the square of the order of the Galois field. Through some simplified methods, such as fast fourier transform, extended min-sum, and min-max, the decoding complexity can be effectively reduced at the expense of some performance. Despite this, \ac{NB-LDPC} codes have exhibited better burst error resistance and superior error-correcting performance at moderate-to-short code lengths, compared to their binary realization. Moreover, \ac{NB-LDPC} codes are particularly well-suited for coded modulation systems, as the coded symbols may be directly mapped to modulated symbols without a bit-to-symbol modulator.

\subsection{SC-LDPC Codes}
The concept of \ac{SC-LDPC} codes originated from convolutional \ac{LDPC} codes proposed in the late 1990s, which inherit the characteristics of convolution and sparse \acp{PCM}. The convolutional structure associates the \acp{CN} of an \ac{LDPC} codeword with the \acp{VN} of the preceding codewords, i.e., the spatial coupling of several LDPC codes. In \cite{Kudekar2011Thres}, the threshold saturation phenomenon of regular \ac{SC-LDPC} codes was discovered for \acp{BEC}, which implies that this class of codes is capacity-achieving under \ac{BP} decoding as the degrees of \acp{VN} and \acp{CN} become sufficiently large, while \ac{MAP} decoding is required for regular \ac{LDPC} codes. Subsequently, this phenomenon was further generalized to binary-input memoryless symmetric channels. The promise of performance has motivated numerous research on \ac{SC-LDPC} codes. It has been shown that the introduction of coupling structure can significantly enhance the performance of waterfall region, outperforming  the \ac{LDPC} codes with the same degree distribution.

\subsection{Polar Codes}
The key idea behind polar codes is the channel polarization. The polarization process begins with the 2$\times$2 Arıkan's polarization kernel (matrix). 
Arıkan showed that the kernel can transform two uses of a \ac{B-DMC} into a good and a bad polarized channel without loss of the overall capacity. Moreover, by recursively applying Arıkan's kernel, the channel polarization of polar codes with a power-of-two code length is achieved. As code length increases, the polarization phenomenon becomes more remarkable, leading to the convergence of the polarized channel capacity to 1 (almost noiseless) or 0 (almost pure-noise). Moreover, the amount of noiseless polarized channels is proved to asymptotically converge to the overall capacity of the original \acp{B-DMC}, i.e. polar codes are capacity-achieving.

Given a message sequence, the general idea of polar encoding is to transmit the message over the most reliable polarized channels, while the remaining polarized channels are filled with frozen bits, which are fixed values known to both the transmitter and the receiver. At the receiver, a \ac{SC} decoder \cite{Arikan2009Channel} is performed to successively output the estimation of message. Moreover, the \ac{SCL} decoder proposed in \cite{Tal2011List} can output a list of candidate codewords and significantly improve decoding performance. As a cost, the complexity of \ac{SCL} decoding is increased by a multiple of the list size compared to \ac{SC} decoding. With \ac{CRC} aided selection of the candidate codewords \cite{Niu2012CRC}, the performance of \ac{SCL} decoding can be further improved.

\subsection{PAC Codes}
Although polar codes have shown to be capacity-achieving, insufficient polarization at short code lengths will cause a capacity loss. That is because the bad polarized channels are not bad enough, and we do not use them for message transmission. To address this issue, Arıkan proposed \ac{PAC} codes in \cite{Arikan2019From}, which adjusts the polarization process by introducing convolutional encoding and sequential decoding.

In \ac{PAC} codes, a rate-1 convolutional code is cascaded before a standard polar code at the encoding side. Then, the encoding of \ac{PAC} codes is performed by rate-profiling, convolutional encoding, and polar encoding, where rate-profiling is the process of placing frozen bits like conventional polar codes. However, the optimal rate-profiling strategy is still unknown due to the concatenation of convolutional codes. During decoding, since the rate-1 convolutional code is equivalent to a tree code under the rate-profiling constraints, Arıkan uses the Fano sequential decoder to perform a tree-based heuristic search, where the path metrics required by the sequential decoder should be updated via repeatedly running an \ac{SC} decoding.

The simulation results in \cite{Arikan2019From} show that \ac{PAC} codes outperform 5G polar codes, and the performance can reach the dispersion bound of the binary-input \ac{AWGN} channel. Therefore, \ac{PAC} codes have been one of the most competitive coding schemes at short code length.

\section{Can Polar Codes Act as the Component Codes of GLDPC Codes?}
\label{Sec:3}
As \ac{LDPC} and polar codes are favored by various latest communication standards, we are interested in exploring next-generation channel coding techniques relying on their existing coding frameworks. Thus, an intuitive idea is whether polar codes can act as the component codes of \ac{GLDPC} codes. Indeed, the crux of this idea revolves around whether polar codes can facilitate an effective \ac{MPA} for decoding, which is tightly related to the error correction performance.

\begin{figure}[!t]
    \centering
    \includegraphics[width=0.48\textwidth]{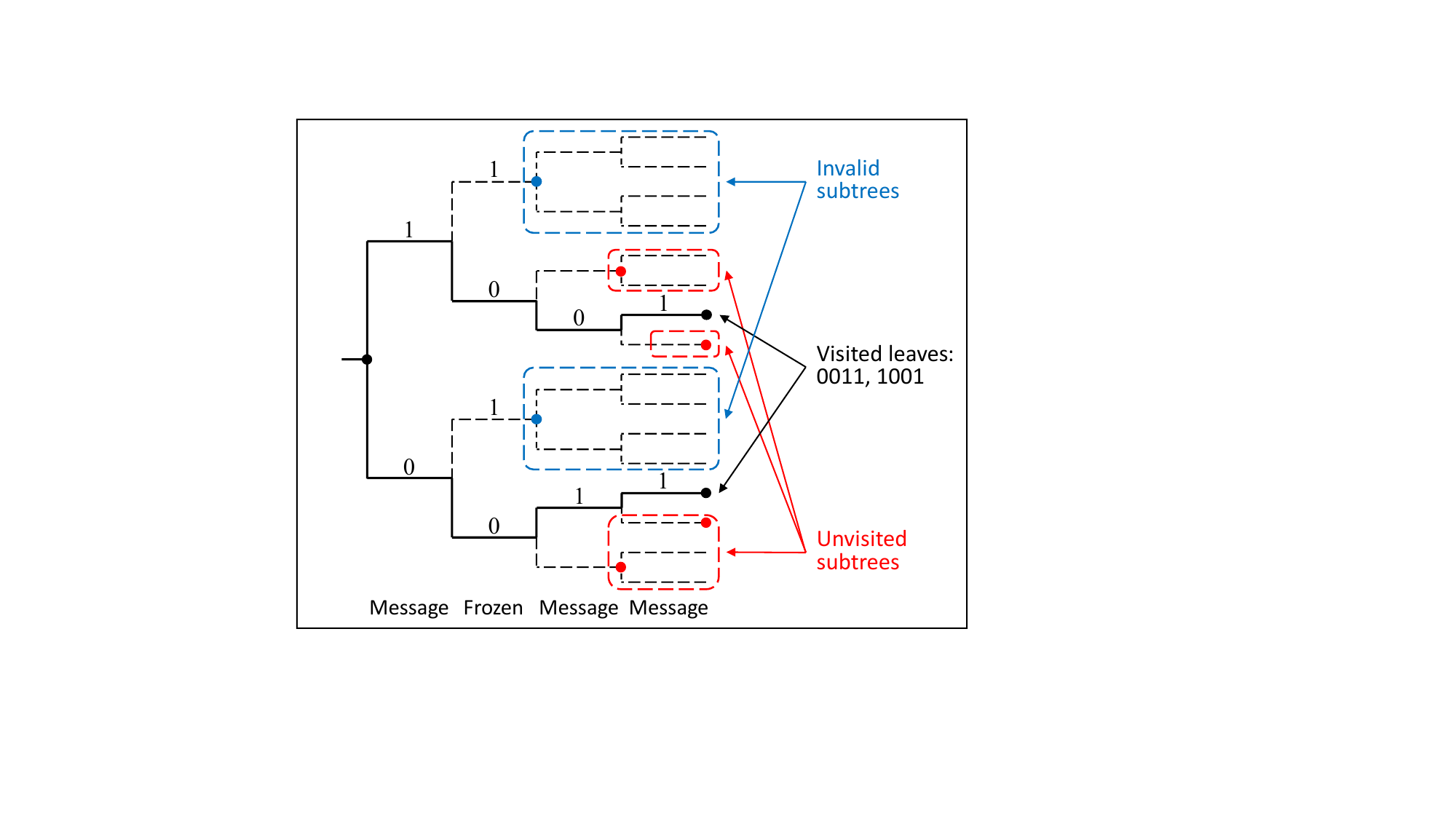}
    \caption{An example of the SCL decoding tree of a polar code with code length of 4, frozen bit equal to 0, and list size of 2 \bl{\cite{Yuan2024Near}}.}
    \label{fig:SCL_eg}
\end{figure}

Generally, the soft information required by \ac{MPA}'s iterations can be extracted from a \ac{SISO} decoder of the component codes. Yet, it is challenging to design a bit-wise \ac{SISO} polar decoder with accurate output and low complexity.

The optimal \ac{SISO} decoder is known to be implemented by the \ac{BCJR} algorithm. However, its complexity grows exponentially with the number of frozen bits, which is unacceptable in practice. Pyndiah's approximation \cite{Pyndiah1998Near} can estimate the \ac{APP} \acp{LLR} from a list of candidate codewords, which has been used for decoding polar product codes in \cite{Coskun2024Prec}. However, Pyndiah's approach assumes that the \acp{APP} of unvisited codewords are directly zero, and as a result, it may output infinite values of \acp{LLR}. Hence, a saturation value is required to avoid unbounded outputs and should be optimized for practical applications as the iterative performance is sensitive to this value. \bl{In \cite{Yuan2023Soft}, a universal \ac{SO-GRAND} without reliance on specific code structures for codes with short-to-moderate redundancy is presented, in which a method that estimates the \acp{APP} omitted by Pyndiah's approach is proposed to provide more accurate soft output.} 

\begin{figure*}[!t]
    \centering
    \includegraphics[width=\textwidth]{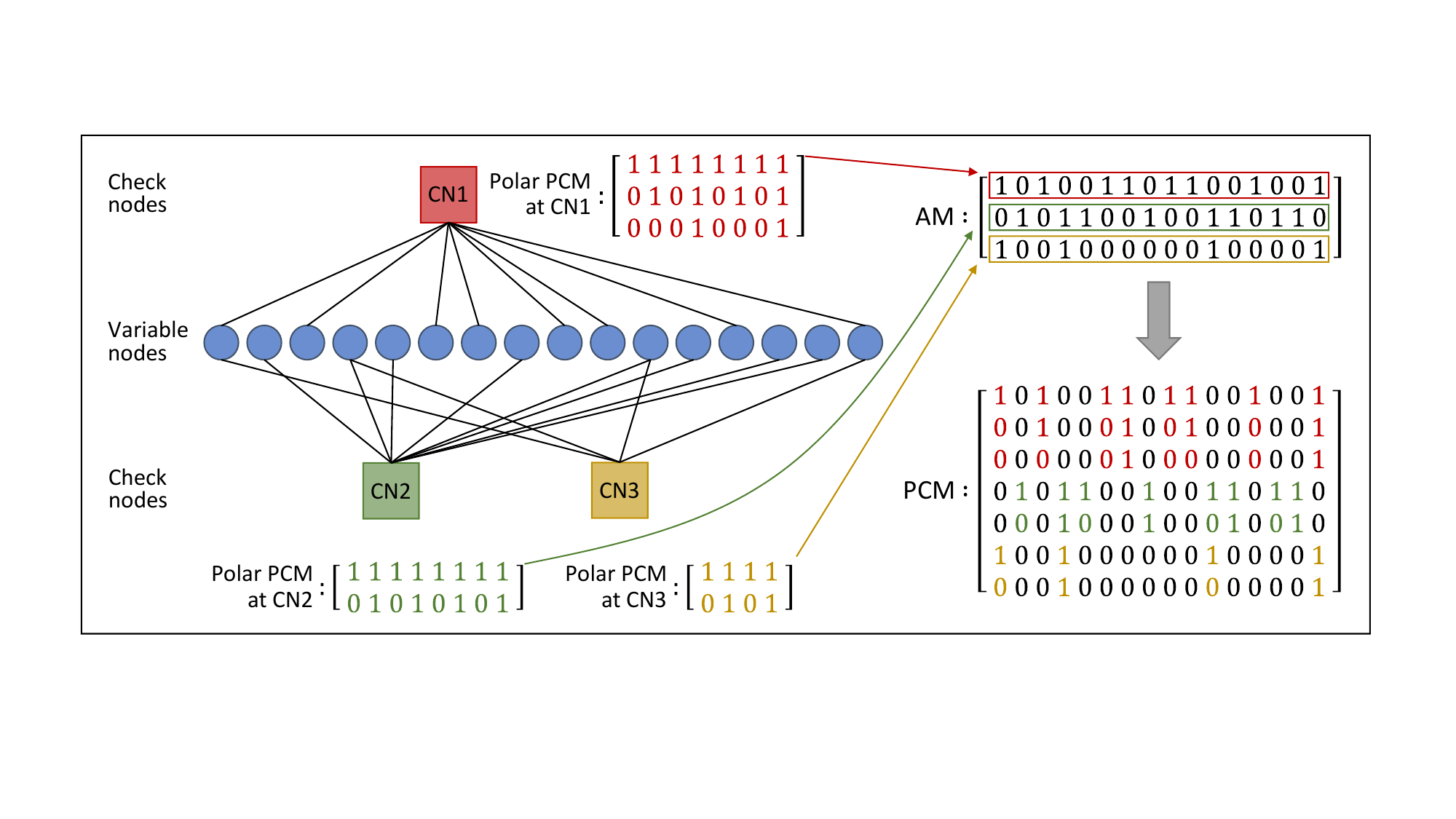}
    \caption{The Tanner graph (left part) and parity check matrix (right part) representations of \ac{GLDPC-PC} codes.}
    \label{fig:GLDPCPC}
\end{figure*}

The above \ac{SISO} decoders are universal and not only applicable to polar codes. Considering the special structure of polar codes, polar \ac{BP} and soft cancellation (SCAN) decoders can provide soft output, while their accuracy is affected by short circles in the polar factor graph and thus inner iterations are required. Cascading an additional \ac{SCL} decoder may improve their performance and avoid iterations. \bl{By extending the methodology in \cite{Yuan2023Soft}, a \ac{SO-SCL} decoder for polar-like codes with comparable complexity to \ac{SCL} decoding was proposed \cite{Yuan2024Near}.} The new approach begins with the \ac{SCL} decoding tree, as shown in Fig. \ref{fig:SCL_eg}, where each leaf represents a possible codeword. After \ac{SCL} decoding, the decoding tree consists of three parts \bl{\cite{Yuan2024Near}}:
\begin{enumerate}
    \item \emph{Visited leaves} are the resulting valid codewords output by the \ac{SCL} decoder.
    \item \emph{Unvisited subtrees} denote the subtrees pruned due to list size limitations and thus are not visited, although their leaves are valid codewords.
    \item \emph{Invalid subtrees} indicate the subtrees rooted at the nodes that are not satisfied the frozen bits constraints, whose leaves are invalid codewords.
\end{enumerate}
The \ac{APP} \acp{LLR} should actually be estimated from the \acp{APP} of all valid leaves, where the \acp{APP} of unvisited leaves are unknown from a conventional \ac{SCL} decoding. To address this, \ac{SO-SCL} approximates the \acp{APP} of unvisited leaves based on visited leaves and nodes, while Pyndiah's approach simply ignores these \acp{APP}. Simulation results in \cite{Yuan2024Near} have demonstrated that the proposed approximation provides more accurate soft information compared to Pyndiah's algorithm and approaches the performance of the optimal \ac{BCJR} algorithm.

\bl{The applications of \ac{SO-GRAND} to \ac{GLDPC} and product decoding presented in \cite{Yuan2023Soft} have highlighted the crucial role of an accurate soft-output component decoder. Later, the emergence of \ac{SO-SCL} \cite{Yuan2024Near} has provided an accurate \ac{SCL}-based soft-output method. Inspired by these works, we believe that polar codes as component codes for \ac{GLDPC} are practical, since the soft information of polar codes can be accurately and efficiently approximated through the widely used \ac{SCL} decoder. Therefore, we adopt \ac{SO-SCL} \cite{Yuan2024Near} for \ac{GLDPC-PC} decoding in our simulations.}

\section{Concepts of GLDPC-PC codes}
Since polar codes have the potential to be the component codes of \ac{GLDPC}, we introduce the primary concepts of such codes in this section, which is referred to as \ac{GLDPC-PC} codes hereafter.

\subsection{Code Structure}
Analogously to \ac{LDPC} codes, \ac{GLDPC-PC} codes can also be represented in the form of Tanner graphs and \acp{PCM}. In the Tanner graph, each \ac{VN} still denotes the corresponding bit in the codeword, while each \ac{CN} constrains the \acp{VN} connected to it to be a polar codeword. As such, the \ac{AM} of the Tanner graph no longer serves as the \ac{PCM} of \ac{GLDPC-PC} codes. Nevertheless, the \ac{PCM} can be drawn from the \ac{AM} by replacing the 1’s in the i-th row of \ac{AM} with the columns of the \ac{PCM} of the i-th component code. \bl{The \ac{PCM} of polar component codes could be obtained through well-known polar construction methods like 5G construction.} Moreover, we remark that, by applying different column permutations to the \ac{PCM} of component codes, the same component code can impose different constraints on the same \acp{VN}.

An example of \ac{GLDPC-PC} codes is illustrated in Fig. \ref{fig:GLDPCPC}. In the Tanner graph (left part of Fig. \ref{fig:GLDPCPC}), each \ac{CN} is constrained by different polar codes, which are distinguished by different colors. The \ac{AM} of the Tanner graph is then written in the upper right corner of Fig. \ref{fig:GLDPCPC}. To obtain the \ac{PCM} of such a code, the k-th 1 in the i-th row of the \ac{AM} is replaced by the k-th column of the permutated \ac{PCM} of the i-th component code, while the remaining positions are filled with all-zero vectors to construct a submatrix. Here, the column permutations are assumed to perform no operation for simplicity. Thus, the resulting matrix is the \ac{PCM} of the illustrated \ac{GLDPC-PC} code, as shown in the bottom right corner of Fig. \ref{fig:GLDPCPC}.

\subsection{Encoding}
A general \ac{GLDPC-PC} encoding method is to multiply the message sequence by the \ac{GM}. Specifically, the \ac{PCM} of \ac{GLDPC-PC} codes can be put into the form of the right submatrix being the identity matrix via Gaussian elimination. Then, the \ac{GM} is directly written from the regained \ac{PCM}. Note that due to the linear combination of the rows of the \ac{PCM}, the resulting \ac{GM} is most likely not sparse, which leads to high computational complexity (depending on the denseness of \ac{GM}) of matrix multiplication when encoding. Nevertheless, with some special code structures, such as the quasi-cyclic form, one can follow the idea in \cite{Li2006Efficient} to perform encoding with lower complexity.

\subsection{Decoding}
The decoding of \ac{GLDPC-PC} codes can be efficiently realized through the well-known \ac{MPA}. A guided process of the \ac{MPA} for \ac{GLDPC-PC} decoding is described as follows.
\begin{enumerate}
    \item Initialize the soft information at \acp{VN} with the channel \acp{LLR} calculated from the received noisy codeword, while all soft information that \acp{CN} pass to \acp{VN} and all soft information that \acp{VN} pass to \acp{CN} are initialized to zero.
    \item \ac{VN} update: At each \ac{VN}, extract the soft information that will be passed to \acp{CN} connected to this \ac{VN} from \ac{LLR} at this \ac{VN} and soft information passed to this \ac{VN} like conventional \ac{LDPC} codes. Repeat this step until all \acp{VN} are updated.
    \item \ac{CN} update: At each \ac{CN}, perform a \ac{SISO} polar decoder with all soft information passed to this \ac{CN} as inputs, and then output soft information that will be passed to \acp{VN} connected to this \ac{CN}. Repeat this step until all \acp{CN} are updated.
    \item Approximate the \ac{APP} \ac{LLR} at each \ac{VN} given \ac{LLR} at this \ac{VN} and soft information passed to this \ac{VN}, and then perform a hard decision on the sequence of \ac{APP} \acp{LLR} to obtain the estimated codeword.
    \item If the estimated codeword fulfills the parity check constraint, or the maximum number of iterations is reached, stop decoding and output the estimate. Otherwise, go back to step 2.
\end{enumerate}

\bl{Note that the \ac{SISO} polar decoder could be efficient schemes like Pyndiah’s approach \cite{Pyndiah1998Near}, \ac{SO-GRAND} \cite{Yuan2023Soft}, and \ac{SO-SCL} \cite{Yuan2024Near} as mentioned in Sec. \ref{Sec:3}.}

\begin{figure}[!t]
    \centering
    \includegraphics[width=0.48\textwidth]{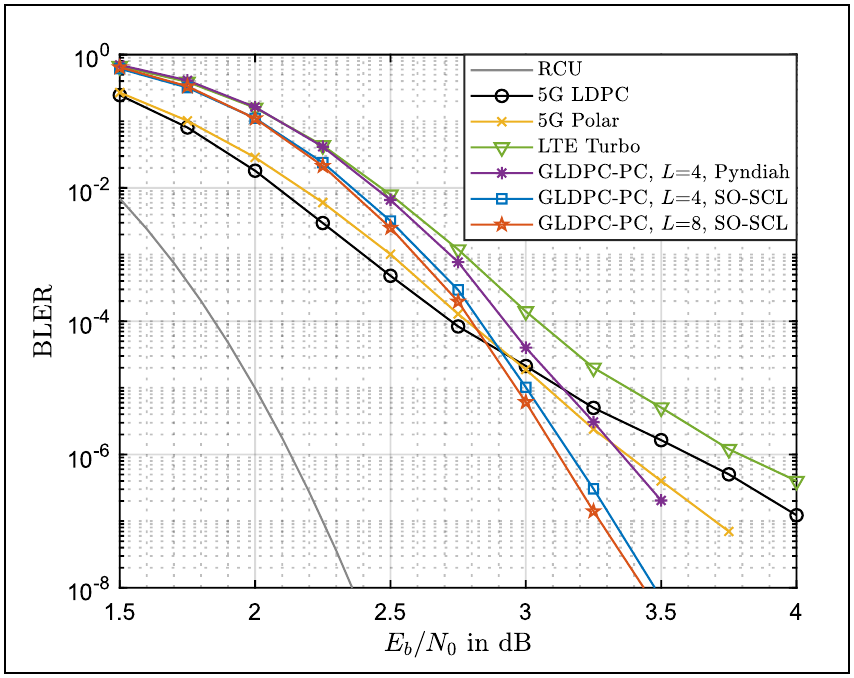}
    \caption{BLER performance of \ac{GLDPC-PC} codes over \ac{AWGN} channel with $N=1024$ and $K=640$.}
    \label{fig:BLER1}
\end{figure}
\begin{figure}[!t]
    \centering
    \includegraphics[width=0.48\textwidth]{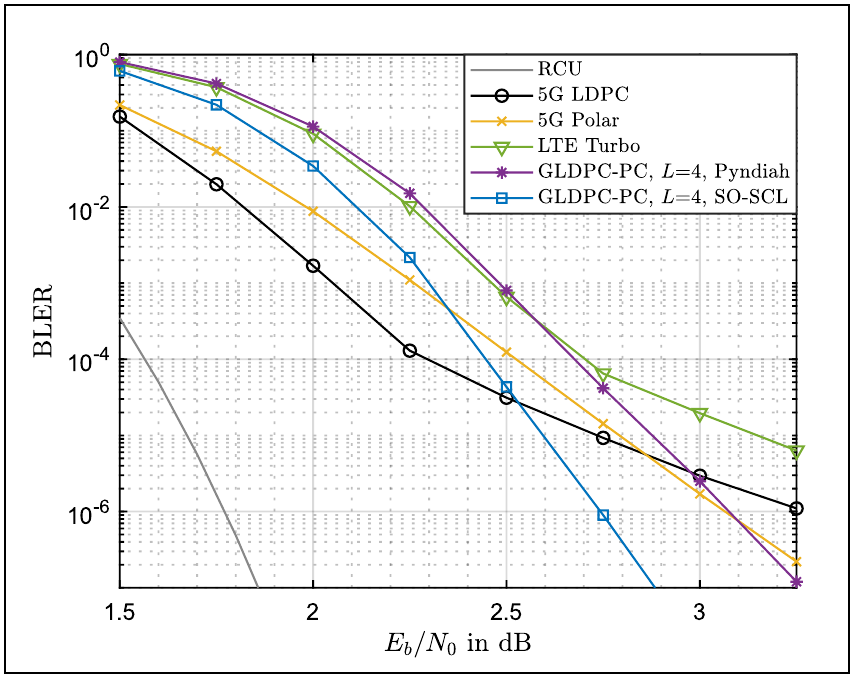}
    \caption{BLER performance of \ac{GLDPC-PC} codes over \ac{AWGN} channel with $N=2048$ and $K=1280$.}
    \label{fig:BLER2}
\end{figure}
\begin{figure}[!t]
    \centering
    \includegraphics[width=0.48\textwidth]{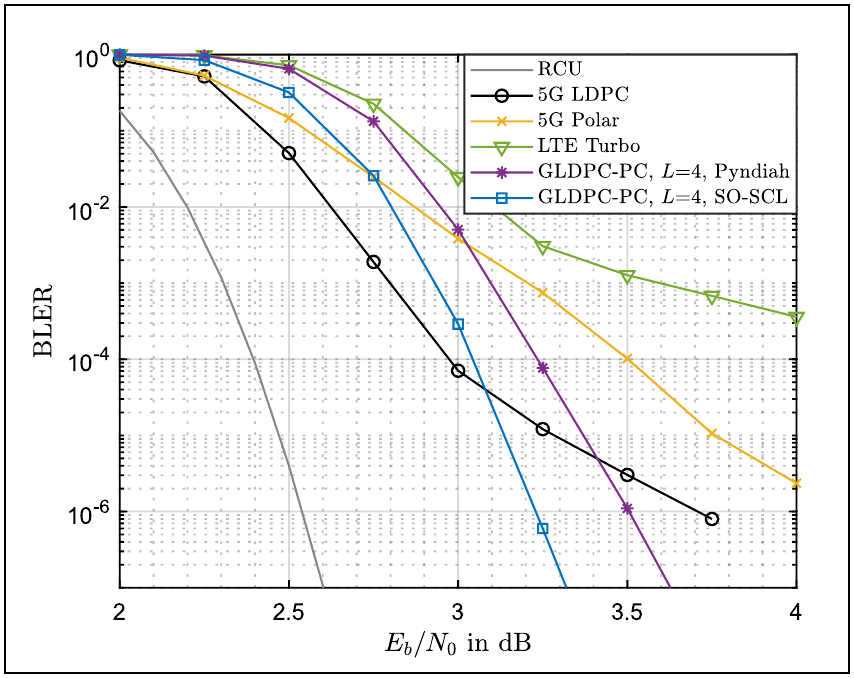}
    \caption{BLER performance of \ac{GLDPC-PC} codes over \ac{AWGN} channel with $N=4096$ and $K=3200$.}
    \label{fig:BLER3}
\end{figure}

\subsection{Performance Evaluation}
Some examples of \ac{GLDPC-PC} codes are offered to illustrate their performance. \bl{We adopt the \ac{GLDPC} structure used in \cite{Yuan2023Soft} to construct \ac{GLDPC-PC} codes, which is acquired from product codes following the idea in \cite[Sec. \MakeUppercase{\romannumeral 5}]{Lentmaier2010From}.} In this method, the \ac{AM} $\bm{\varGamma}$ is represented as
\begin{equation}
    \bm{\varGamma} = \left[\begin{matrix}
    \bm{I}^0 &\bm{I}^0 &\bm{I}^0 &\cdots &\bm{I}^{0~~~}\\
    \bm{I}^0 &\bm{I}^1 &\bm{I}^2 &\cdots &\bm{I}^{n-1}
    \end{matrix}\right],
\end{equation}
where $\bm{I}^i$ is the $i$-th cyclic right shift of the $\frac{M}{2} \times \frac{M}{2}$ identity matrix $\bm{I}^0$ and $M$ is the number of \acp{CN}. The first and second “block” row in $\bm{\varGamma}$ are associated to a $(n, k)$ polar component code and the code obtained by random column permutations of its \ac{PCM}, respectively. Thus, the resulting code length $N$ of \ac{GLDPC-PC} codes is $n^2$ and parity check number is $M(n-k)$. Note that the information length $K$ may be greater than $n^2-M(n-k)$ if some parity checks are linearly dependent. The (1024, 640), (2048, 1280), and (4096, 3200) \ac{GLDPC-PC} codes are constructed from (32, 26) polar code with $M=64$, (32, 26) polar code with $M=128$, and (64, 57) polar code with $M=128$, respectively.

Figs. \ref{fig:BLER1}, \ref{fig:BLER2}, and \ref{fig:BLER3} present the \ac{BLER} performance as a function of energy per bit vs. power spectral density of noise ($E_b/N_0$) for \ac{GLDPC-PC} codes over \ac{AWGN} channel with different code length and rate. We also provide the results of 5G \ac{LDPC} codes, 5G polar codes, and LTE turbo codes for comparison. The \ac{RCU} bound is simultaneously displayed. \bl{We use Pyndiah’s approach \cite{Pyndiah1998Near} and \ac{SO-SCL} \cite{Yuan2024Near} for \ac{GLDPC-PC} decoding, and} the maximum number of iterations for \ac{GLDPC-PC} decoding is 20. We adopt \ac{BP} algorithm with a maximum of 50 iterations for 5G \ac{LDPC} codes and \ac{CRC}-aided \ac{SCL} decoder with 16-bit \ac{CRC} and list size 16 for 5G polar codes. We observe that \ac{GLDPC-PC} codes outperform other codes at moderate-to-high \acp{SNR}. Specifically, \ac{GLDPC-PC} codes with \ac{SO-SCL} achieve positive gain at a \ac{BLER} of $10^{-4}\sim 10^{-5}$ compared to 5G \ac{LDPC} codes. This gain is reduced by applying the inaccurate approximation of Pyndiah's approach. Whereas the \ac{BLER} performance of \ac{GLDPC-PC} codes can be further improved by increasing the list size. Moreover, \ac{GLDPC-PC} codes show no apparent error floor at least when \ac{BLER} drops to $10^{-7}$, while the turning point of the waterfall region for \ac{LDPC} codes is at a \ac{BLER} of about $10^{-4}$.

\begin{figure}[!t]
    \centering
    \includegraphics[width=0.48\textwidth]{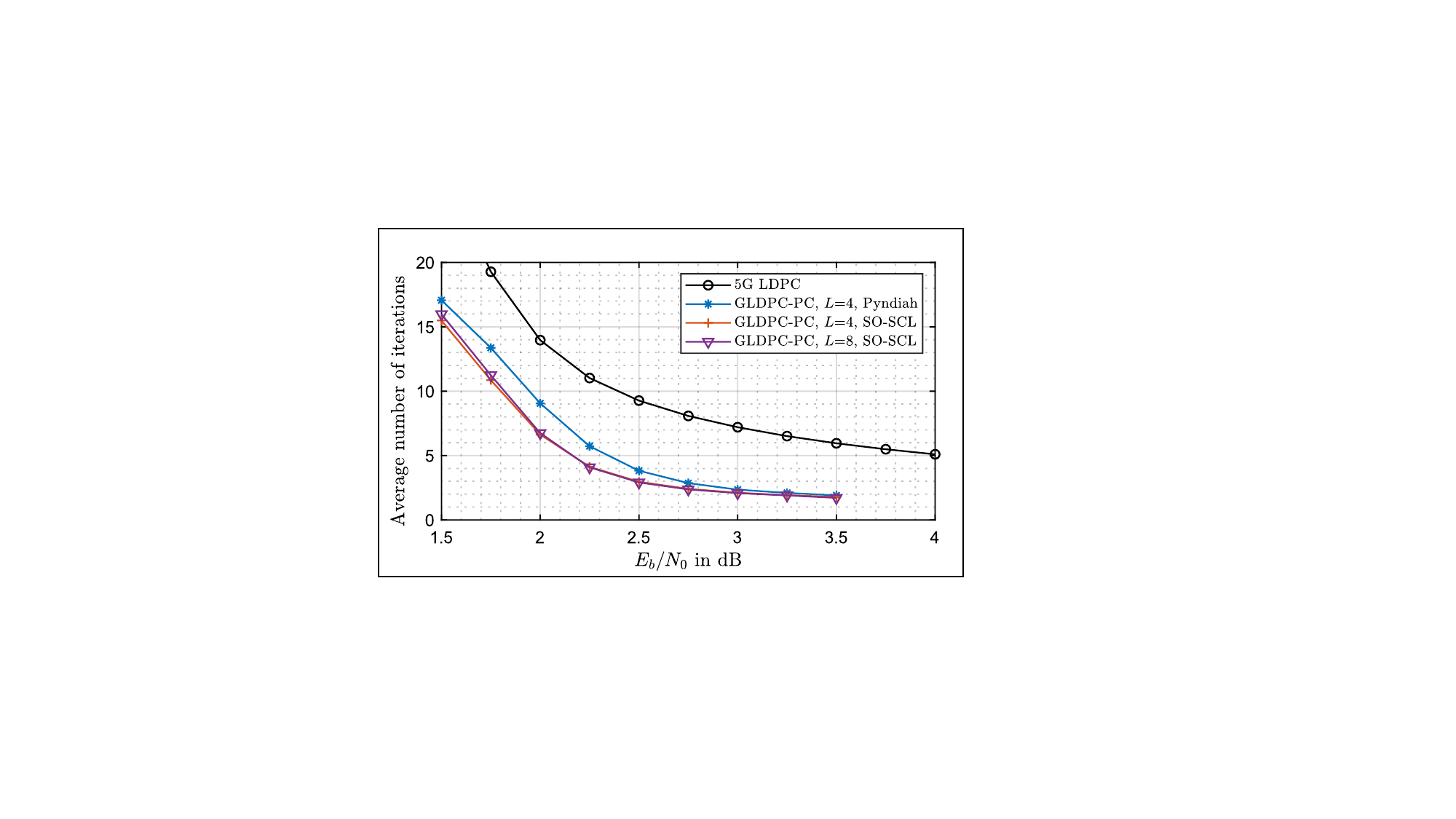}
    \caption{Average number of iterations for decoding \ac{GLDPC-PC} codes and 5G \ac{LDPC} codes, with $N = 1024$ and $K=640$.}
    \label{fig:Iavg}
\end{figure}

Assuming that $I_{\rm avg}$ is the average number of iterations, $M$ is the number of \acp{CN}, $N$ is the code length, and $\bar{d}_v$($\bar{d}_c$) denotes the average degree distribution of \acp{VN}(\acp{CN}), the \ac{BP} decoder of \ac{LDPC} codes performs 
\begin{equation}
    I_{\rm avg}(N\bar{d}_v+M\bar{d}_c)
\end{equation}
basic operations for one received codeword. By replacing the single parity check with polar codes, the \ac{GLDPC-PC} decoder executes about 
\begin{equation}
    I_{\rm avg}(N\bar{d}_v+L\sum_{i=1}^M n_i(\log n_i+1))
\end{equation}
basic operations, where $L$ is the list size and $n_i$ is the code length of the $i$-th component code. Since polar codes impose stronger constraints on \acp{CN}, the average number of iterations of \ac{GLDPC-PC} codes is approximately one-third to half of that of 5G \ac{LDPC} codes, as verified in Fig. \ref{fig:Iavg}. Thus, in our simulation settings, the decoding computational complexity of \ac{GLDPC-PC} codes with $L=4$ is roughly two to three times that of 5G \ac{LDPC} codes and comparable to that of polar codes. Please note that the presented \ac{GLDPC-PC} codes are obtained from off-the-shelf structures and not specifically designed for polar component. A general code construction method remains to be investigated. Moreover, the decoding complexity could be further reduced by applying a low-complexity \ac{SCL} decoder, e.g. node-based \ac{SCL}, with which it is expected to achieve a complexity comparable to 5G \ac{LDPC} codes.

\section{Opportunities and Challenges}
\ac{GLDPC-PC} codes achieve a unity of \ac{LDPC} codes and polar codes, and demonstrate promising performance on error correction and complexity. In this section, we highlight the superiorities and challenges of applying \ac{GLDPC-PC} codes to 6G communication systems and beyond.

\subsection{Superiorities for Practical Application}
\subsubsection{Ultra-Reliability}
In contrast to 5G \ac{LDPC} codes that have error floors at a \ac{BLER} of $10^{-4}\!\sim\!10^{-5}$, \ac{GLDPC-PC} codes have shown the potential to enable waterfall region down to $10^{-8}$ or below while achieving gains. This is in line with the demand for future 6G ultra-reliable communications.
\subsubsection{Versatile Decoder}
Since the \ac{SO-SCL} decoding relies on the process of hard-output \ac{SCL} decoding, we possess the flexibility to configure the \ac{SO-SCL} decoder to enable decoding for either conventional polar codes or component codes of \ac{GLDPC-PC} codes.
\subsubsection{Parallel Decoding}
To reduce the decoding latency, parallel decoding is widely used in hardware implementation of conventional LDPC codes. Similarly, \ac{GLDPC-PC} codes can achieve parallel update of \acp{CN} by executing multiple polar \ac{SISO} decoders simultaneously, which will greatly reduce the decoding latency.
\subsubsection{Diverse application scenarios}
\Ac{MIMO}, fiber-optic, and other systems can also utilize \ac{GLDPC-PC} codes to pursue more reliable communication. Particularly, with iterative detection and decoding, \ac{GLDPC-PC} codes will further contribute to the error correction, multi-user access, and code-aided channel estimation performance in \ac{MIMO} systems.

\subsection{Challenges and Future Directions}
\subsubsection{Code Structure Design}
The code structure design of \ac{GLDPC-PC} codes involves which \acp{VN} are connected to one \ac{CN} and which polar component code each \ac{CN} is constrained by. The structure is related to the \ac{BLER} performance (including waterfall region and error floors), the encoding complexity and the decoding complexity of \ac{GLDPC-PC} codes. In general, there will be a trade-off between these technical indicators. Yet, how to design a good \ac{GLDPC-PC} code is unknown and remains to be studied. Moreover, structured \ac{GLDPC-PC} codes that support rate compatibility and hybrid automatic repeat request should also be considered.
\subsubsection{Encoding Complexity}
Although it was mentioned in the previous section that \ac{GLDPC-PC} codes with quasi-cyclic structure can achieve linear-time encoding, we would like to develop more generalized low-complexity encoding schemes. One possible direction is to explore the utilization of polar transform for efficient encoding.
\subsubsection{Decoding Complexity}
The decoding complexity of \ac{GLDPC-PC} codes is mainly affected by the density of \acp{CN}, the average degree distribution of \acp{CN}, and the complexity of \ac{SISO} polar decoder. The first two factors are considered during the code structure design, and there is a trade-off that fewer \acp{CN} and lower average degree of help to reduce the decoding complexity, but will inevitably cause the loss of error correction performance. As for the complexity reduction of \ac{SISO} polar decoding, it is an ongoing topic in polar codes research.
\subsubsection{Decoding Latency}
As an important technical indicator for hardware implementation, decoding latency is not necessarily proportional to decoding complexity, as it is also related to factors such as parallelism, cache size, and memory access times. However, the decoding latency of \ac{GLDPC-PC} codes has not been comprehensively evaluated, which is a direction for future work.

\section{Conclusion}
This article focuses on channel coding towards \ac{6G} communications, including an overview of state-of-the-art channel codes. Generally, to avoid significant modifications to existing standards, improvements based on well-established coding frameworks like \ac{LDPC} and polar codes are more readily acceptable. Thus, we specifically discuss imposing stronger \ac{CN} constraints in \ac{LDPC} codes by utilizing polar codes, namely \ac{GLDPC-PC} codes, \bl{where the soft information passed by polar components to \acp{VN} is efficiently extracted from the \ac{SO-SCL} decoder \cite{Yuan2024Near}}. \ac{GLDPC-PC} codes have demonstrated promising \ac{BLER} performance, especially in terms of ultra-low error floor, which are compatible with the demand for higher data rate and ultra-reliable transmission of 6G communications. Finally, we raise some relevant open issues regarding the practical application of \ac{GLDPC-PC} codes.


\section*{Biographies}
\footnotesize
\noindent\textbf{Li Shen}
received B.S. and M.S. degrees from Shanghai Jiao Tong University, China, in 2021 and 2024, respectively. He is currently pursuing a Ph.D. degree at the Department of Electronic Engineering, Shanghai Jiao Tong University. His research interests are channel coding and coded modulation.

\vspace{\baselineskip}
\noindent\textbf{Yongpeng Wu} [SM]
received the B.S. degree in telecommunication engineering from Wuhan University, China, in 2007, and the Ph.D. degree in Southeast University, China, in 2013. Dr. Wu is currently a Professor with the Department of Electronic Engineering, Shanghai Jiao Tong University, China. His research interests include massive MIMO/MIMO systems, massive machine type communication, physical layer security, and signal processing for wireless communications.

\vspace{\baselineskip}
\noindent\textbf{Yin Xu} [M]
received the B.Sc. degree in information science and engineering from Southeast University,	China, in 2009, and the M.S. and Ph.D. degrees in electronics engineering from Shanghai Jiao Tong University, in 2011 and 2015, respectively, where he currently works as an Assistant Professor. His main research interests include channel coding, advanced bit-interleaved coded modulation and other physical layer technologies in broadcasting and 5G.

\vspace{\baselineskip}
\noindent\textbf{Xiaohu You} [F]
received his Master and Ph.D. Degrees from Southeast University, China, in Electrical Engineering in 1985 and 1988, respectively. Since 1990, he has been working with National Mobile Communications Research Laboratory at Southeast University, where he is currently professor and director of the Lab. His current research interests include wireless networks, advanced signal processing and its applications.

\vspace{\baselineskip}
\noindent\textbf{Xiqi Gao} [F]
received his Ph.D. degree in electrical engineering from Southeast University, China, in 1997. He joined the Department of Radio Engineering, Southeast University, in 1992. Since 2001, he has been a professor of information systems and communications. His current research interests include broadband multi-carrier communications, massive MIMO wireless communications, satellite communications, optical wireless communications, information theory and signal processing for wireless communications.

\vspace{\baselineskip}
\noindent\textbf{Wenjun Zhang} [F]
received the B.S., M.S., and Ph.D. degrees in electronic engineering from Shanghai Jiao Tong University, China, in 1984, 1987, and 1989, respectively. After three years’ working as an Engineer with Philips, Germany, he went back to his Alma Mater in 1993, and became a Full Professor of Electronic Engineering in 1995. His main research interests include video coding and wireless transmission, multimedia semantic analysis, and broadcast/broadband network convergence.


\begin{thebibliography}{15}
\bibliographystyle{IEEEtran}
\bibitem{Gallager1962Low}
R. Gallager, “Low-Density Parity-Check Codes,” \emph{{IRE} Trans. Info. Theory}, vol. 7, no. 1, Jan. 1962, pp. 21–28.

\bibitem{Arikan2009Channel}
E. Arıkan, “Channel Polarization: A Method for Constructing Capacity-Achieving Codes for Symmetric Binary-Input Memoryless Channels,” \emph{{IEEE} Trans. Inf. Theory}, vol. 55, no. 7, Jul. 2009, pp. 3051–3073.

\bibitem{Rowshan2024Channel}
M. Rowshan \emph{et al.}, “Channel Coding Toward 6G: Technical Overview and Outlook,” \emph{{IEEE} Open J. Commun. Soc.}, vol. 5, Apr. 2024, pp. 2585–2685.

\bibitem{Tanner1981Recur}
R. Tanner, “A Recursive Approach to Low Complexity Codes,” \emph{{IEEE} Trans. Inf. Theory}, vol. IT-27, no. 5, Sep. 1981, pp. 533–547.

\bibitem{Davey1998Low}
M.C. Davey and D. MacKay, “Low-Density Parity Check Codes over GF(q),” in \emph{Proc. Inf. Theory Workshop (ITW)}, Jun. 1998, pp. 70–71.

\bibitem{Kudekar2011Thres}
S. Kudekar, T. J. Richardson, and R. L. Urbanke, “Threshold Saturation via Spatial Coupling: Why Convolutional LDPC Ensembles Perform So Well over the BEC,” \emph{{IEEE} Trans. Inf. Theory}, vol. 57, no. 2, Jan. 2011, pp. 803–834.

\bibitem{Arikan2019From}
E. Arıkan, “From Sequential Decoding to Channel Polarization and Back Again,” \emph{arXiv preprint arXiv:1908.09594v3}, Sep. 2019.


\bibitem{Coskun2024Prec}
M. C. Co\c{s}kun, “Precoded Polar Product Codes,” in \emph{Proc. {IEEE} Int. Symp. Inf. Theory (ISIT)}, Jul. 2024, pp. 3368–3733.

\bibitem{Tal2011List}
I. Tal and A. Vardy, “List Decoding of Polar Codes,” \emph{{IEEE} Trans. Inf. Theory}, vol. 61, no. 5, May 2015, pp. 2213–2226.

\bibitem{Niu2012CRC}
K. Niu and K. Chen, “CRC-Aided Decoding of Polar Codes,” \emph{{IEEE} Commun. Lett.}, vol. 16, no. 10, Oct. 2012, pp. 1668–71.


\bibitem{Pyndiah1998Near}
R. Pyndiah, “Near-Optimum Decoding of Product Codes: Block Turbo Codes,” \emph{{IEEE} Trans. Commun.}, vol. 46, no. 8, Aug. 1998, pp. 1003–1010.

\bibitem{Yuan2023Soft}
P. Yuan \emph{et al.}, “Soft-output (SO) GRAND and Iterative Decoding to Outperform LDPC Codes,” \emph{{IEEE} Trans. Wireless Commun. (Early Access)}, Jan. 2025, pp. 1-14.

\bibitem{Yuan2024Near}
P. Yuan, K. R. Duffy, and M. Médard, “Soft-Output Successive Cancellation List Decoding,” \emph{{IEEE} Trans. Inf. Theory}, vol. 71, no. 2, Feb. 2025, pp. 1007-1017.

\bibitem{Li2006Efficient}
Z. Li \emph{et al.}, “Efficient Encoding of Quasi-Cyclic Low-Density Parity-Check Codes,” \emph{{IEEE} Trans. Commun.}, vol. 54, no. 1, Jan. 2006, pp. 71–81.

\bibitem{Lentmaier2010From}
M. Lentmaier \emph{et al.}, “From Product Codes to Structured Generalized LDPC Codes,” in \emph{Proc. 5th Int. ICST Conf. Commun. Netw. China (CHINACOM)}, Aug. 2010, pp. 1–8.

\end{thebibliography}
\end{document}